\documentclass[manuscript]{aastex}
\usepackage{emulateapj5}

\newcommand{\nuc}[2]{${}^{#2} \rm #1$}

\def\gtaprx {\lower .14ex\hbox{\rlap{\raise .9ex\hbox{\hskip .3ex
	{\ifmmode{\scriptscriptstyle >}\else
		{$\scriptscriptstyle >$}\fi}}}
	\kern -.4ex{\ifmmode{\scriptscriptstyle \sim}\else
		{$\scriptscriptstyle\sim$}\fi}}}
\def\ltaprx {\lower .14ex\hbox{\rlap{\raise .9ex\hbox{\hskip .3ex
	{\ifmmode{\scriptscriptstyle <}\else
		{$\scriptscriptstyle <$}\fi}}}
	\kern -.4ex{\ifmmode{\scriptscriptstyle \sim}\else
		{$\scriptscriptstyle\sim$}\fi}}}

\shorttitle{P-Process inside SN-Driven Supercritical Disks}
\shortauthors{Fujimoto et al.}

\begin{document}

\title{
P-Process Nucleosynthesis 
inside Supernova-Driven Supercritical Accretion Disks
}

\author{Shin-ichirou Fujimoto}
\affil{Department of Electronic Control, 
Kumamoto National College of Technology, Kumamoto 861-1102, Japan;
fujimoto@ec.knct.ac.jp}

\author{Masa-aki Hashimoto and Osamu Koike}
\affil{Department of Physics, School of Sciences, 
Kyushu University, Fukuoka 810-8560, Japan;
hashi@gemini.rc.kyushu-u.ac.jp, koike@gemini.rc.kyushu-u.ac.jp} 

\and

\author{Kenzo Arai and Ryuichi Matsuba}
\affil{Department of Physics, Kumamoto University, Kumamoto 860-8555, Japan;
arai@sci.kumamoto-u.ac.jp, matsuba@sci.kumamoto-u.ac.jp}


\begin{abstract}
 We investigate p-process nucleosynthesis in a supercritical accretion disk
 around a compact object of 1.4 $M_{\odot}$, 
 using the self-similar solution of an optically thick 
 advection dominated flow. 
 Supercritical accretion is expected to occur in a supernova with 
 fallback material accreting onto a new-born compact object. 
 It is found that appreciable amounts of p-nuclei are synthesized 
 via the p-process in supernova-driven supercritical accretion disks (SSADs)
 when the accretion rate $\dot{m} = \dot{M}c^2/(16 L_{\rm Edd}) >10^5$, 
 where $L_{\rm Edd}$ is the Eddington luminosity. 
 Abundance profiles of p-nuclei ejected from SSADs
 have similar feature to those of the oxygen/neon layers 
 in Type II supernovae when the abundance of 
 the fallback gas far from the compact object is that of the oxygen/neon
 layers in the progenitor.
 The overall abundance profile is in agreement with that of the solar system.
 Some p-nuclei, such as Mo, Ru, Sn, and La,  
 are underproduced in the SSADs as in Type II supernovae.
 If the fallback gas is mixed with a small fraction of proton
 through Rayleigh-Taylor instability during the explosion, 
 significant amounts of \nuc{Mo}{92} are produced inside the SSADs.
 \nuc{Ru}{96} and \nuc{La}{138} are also produced 
 when the fallback gas contains abundant proton
 though the overall abundance profile of p-nuclei is rather different from 
 that of the solar system. 
The p-process nucleosynthesis in SSADs contributes
to chemical evolution of p-nuclei, in particular \nuc{Mo}{92}, 
if several percents of fallback matter are ejected via jets and/or winds.
\end{abstract}


\keywords{Accretion, accretion disks  --- 
nuclear reactions, nucleosynthesis, abundances --- 
stars: supernovae: general }

\section{Introduction}

There exist 35 neutron deficient stable nuclei with mass number 
$\rm A \ge 74$ referred to as p-nuclei. 
These nuclei are found in meteorites  only inside the solar system; isotopic anomalies
involving these nuclei are also found in some primitive meteorites.
Many production sites of the p-nuclei have been proposed: 
oxygen/neon layers of highly evolved massive stars 
during their presupernova phase (Arnould 1976) and
during their supernova explosion (Woosley \& Howard 1978);
X-ray novae (Schatz et al.~1998);
neutrino-driven winds originated from a nascent neutron star 
shortly after supernova explosion (Hoffman et al.~1996);
Type Ia supernova explosion (Howard, Meyer \& Woosley 1991);
helium-accreting CO white dwarfs of sub-Chandrasekhar mass 
 (Goriely et al.~2002).
Among them, the most promising
is the oxygen/neon layers during Type II supernova explosion.
The p-nuclei are synthesized by the photodisintegrations 
of s-nuclei (s-process seeds) produced in the layers
during the core helium burning in the progenitor.
Photodisintegrate ($\gamma, \rm n$) reactions are followed by
($\gamma,\rm p$) and/or ($\gamma, \alpha$) reactions.
The production of p-nuclei via the subsequent photodisintegrations
is referred to as a p-process.
Extensive investigations of the p-process in Type II supernovae
are performed by Prantzos et al.~(1990b), 
Rayet, Prantzos \& Arnould (1990), Rayet et al.~(1995), and
Arnould, Rayet \& Hashimoto (1998)
with a large nuclear network.
A p-process model of Rayet et al.~(1995) in the oxygen/neon layers 
during Type II supernovae reproduces the overall abundance profile
of p-nuclei in the solar system.
However, the model has a conspicuous shortcoming of underproduction of 
${}^{92,94}\rm Mo$, ${}^{96,98}\rm Ru$, and $^{138}\rm La$.
Abundant productions of such p-nuclei
have been invoked to neutrino processes
(Woosley et al.~1990; Goriely et al.~2001); 
Yield of ${}^{138}\rm La$ in the oxygen/neon layers is increased
via neutrino interactions 
though those of Mo and Ru are not appreciably changed (Goriely et al.~2001).
Enhancement of s-process seeds synthesized in the layers 
during core helium burning could improve 
the underproduction of Mo and Ru with an extremely high rate
of $\rm {}^{22}Na (\alpha, n) {}^{25}Mg$ reaction (Costa et al.~2000).
However, such a high rate is possibly excluded 
by a recent experiment (Jaeger et al.~2001).

During supernova explosion in massive stars, 
certain fraction of ejected matter is expected to 
fall back onto a new-born compact object
with supercritical accretion rates greater than $10^4 \dot{M}_{\rm cr}$
(Woosley \& Weaver 1995, hereafter WW95), 
where $\dot{M}_{\rm cr}$ is the critical mass accretion rate
given by $16 L_{\rm Edd}/c^2$ with
the Eddington luminosity, $L_{\rm Edd}$.
If the fallback material has substantial angular momentum, 
formation of an accretion disk is inevitable 
 (Mineshige et al.~1997).
Fujimoto et al.~(2001) have investigated 
nucleosynthesis inside supernova-driven supercritical accretion disks
 (SSADs)
and shown that appreciable nuclear reactions take place 
if the accretion rate $\dot{M} > 10^5 \dot{M}_{cr}$.
At such high accretion rates, 
SSADs contain the regions where density,
temperature, and dynamical time-scale are comparable
to those of the oxygen/neon layers in Type II supernovae.
Therefore, the p-process is likely to operate efficiently 
inside SSADs.

We propose a SSAD
as a candidate for the production of p-nuclei.
In \S 2, we present a model of SSADs and
a nuclear network to calculate the abundance
evolution inside a SSAD.
We also discuss possible progenitors for p-process nucleosynthesis
inside SSADs and the initial chemical abundances of the
fallback material far from the central compact remnant.
In \S 3, we describe 
abundance distributions of p-nuclei inside SSADs,
ejected masses of these nuclei from SSADs, and
effects of proton and helium contamination of the fallback gas
via Rayleigh-Taylor instability.
We discuss our results in \S 4 
and finally present concluding remarks in \S 5.

\section{Model and Input Physics}

\subsection{The Supercritical Accretion Disk Model}

Physical quantities of supercritical accretion disks 
with $\dot{M} \gg \dot{M}_{\rm cr}$
are well described in terms of the self-similar solution. 
The density $\rho$, temperature $T$ 
and drift time-scale $t_{\rm dr}$ of accreting gas are evaluated 
with the self-similar disk model as adopted by Fujimoto et al.~(2001) 
to be 
\begin{eqnarray}
   \left( \frac{\rho}{10^4 {\rm \, g \, cm^{-3}}} \right)
   &=&
   1.9 (1 + \alpha_{\rm vis}^2/7) 
   \left( \frac{M}{1.4 M_{\odot}} \right)^{-1} \nonumber \\
   &\times&
   \left( \frac{\alpha_{\rm vis}}{0.01} \right)^{-1} 
   \left( \frac{\dot{m}}{10^6} \right)
   \left( \frac{r}{3 r_{\rm g}} \right)^{-3/2}, \\
  \left( \frac{T}{10^9 {\rm \, K}} \right)
   &=& 
   4.3 \left( \frac{M}{1.4 M_{\odot}} \right)^{-1/4} \nonumber \\
   &\times&
   \left( \frac{\alpha_{\rm vis}}{0.01} \right)^{-1/4} 
   \left( \frac{\dot{m}}{10^6} \right)^{1/4} 
   \left( \frac{r}{3r_{\rm g}} \right)^{-5/8}, \\
   \left( \frac{t_{\rm dr}}{0.01 {\rm \, s} } \right) 
   &=& 
   1.9 \sqrt{1 + \alpha_{\rm vis}^2/7} 
   \left( \frac{M}{1.4 M_{\odot}} \right) \nonumber \\
   &\times&
   \left( \frac{\alpha_{\rm vis}}{0.01} \right)^{-1} 
   \left( \frac{r}{3r_{\rm g}} \right)^{3/2},
\end{eqnarray}
where 
$r$ is the radial coordinate, 
$M$ is the mass of the central compact object, 
$\alpha_{\rm vis}$ is the viscous parameter, 
$\dot{m} = \dot{M}/\dot{M}_{\rm cr}$, and 
$r_{\rm g} = 2GM/c^2$ is the Schwarzschild radius.
The model has been assumed to be steady and axisymmetric
with the self-gravity ignored.
It is noted that 
in SSADs with $\dot{m} \ltaprx 10^{12}$ 
advective cooling dominates over radiative and neutrino cooling;
it balances with viscous energy generation.
A disk is also assumed to be extended from 
the last stable circular orbit around a black hole 
to the location of accretion shock $\sim 10^9 \rm cm$
(Fryer, Colgate \& Pinto 1999; MacFadyen, Woosley \& Heger 2001).
Hence, the inner and outer boundaries of the disk, 
$r_{\rm in}$ and $r_{\rm out}$, are set to be 
3 $r_{\rm g}$ and $3 \times 10^3\; r_{\rm g}$ (= 1.2 $\times 10^9 M/M_\odot \rm \, cm$), 
respectively.
A sequence of the models is specified in terms of 
$M$, $\alpha_{\rm vis}$, and $\dot{m}$.
In the present paper, we fix both $M =1.4\; M_{\odot}$ and 
$\alpha_{\rm vis}= 0.01$ (e.g, Hawley 2000), while
$\dot{m}$ is left to be a parameter.

\subsection{The Nuclear Reaction Network}

We have developed a nuclear reaction network
which has been extended from a smaller network 
(Hashimoto \& Arai 1985; Koike et al.~1999).
The nuclear data are taken from 
the data base REACLIB\footnote{
The REACLIB compilation of reaction rates is available at: 
\url{ftp://quasar.physik.unibas.ch/pub/tommy/astro/reaclib/}
}, 
which includes many experimental rates and
theoretical Hauser-Feshbach rates (Rauscher \& Thielemann 2000, 2001).
We have also used the experimental mass data (Audi \& Wapstra 1995).
The reactions included in the network are 
$(\rm n, \gamma)$, $(\rm n, p)$, $(\rm n, \alpha)$, 
$(\rm p, \gamma)$, $(\rm p, \alpha)$, $(\alpha, \gamma)$, 
$3\alpha$, 
and their inverse reactions.
The various channels of ${}^{12}\rm C +{}^{12}\rm C$, 
${}^{12}\rm C +{}^{16}\rm O$, and 
${}^{16}\rm O +{}^{16}\rm O$ are also included in the network.
For $\beta^-$ and $\beta^+$ decays, 
REACLIB rates are updated with the corresponding data 
in Horiguchi et al.~(1996) if available.
We have also added $\beta^-$ and $\beta^+$ decays and 
electron and positron captures (Fuller et al.~1980, 1982a, 1982b).
Some rates are revised with experimental rates
(see details in Koike et al.~1999).
Our network
includes 1988 nuclides from neutron and proton up to 
$\rm {}^{209}Bi$, presented in Table 1.
The network is solved implicitly with an inverter of a sparse matrix
(Timmes 1999).


\subsection{Possible Progenitors for P-Process Nucleosynthesis inside SSADs}

As the case of Type II supernovae, 
synthesis of p-nuclei inside SSADs requires the s-process seeds.
The fallback after a supernova explosion is likely to be induced by 
deep gravitational potential of relatively massive stars and/or 
reverse shock inwardly propagating from the outer composition interfaces 
(Colgate 1971; Chevalier 1989; WW95).
A total amount of fallback material increases as mass of a progenitor are enhanced
(WW95; Fryer 1999, hereafter F99; Fryer \& Heger 2000);
Massive fallback is likely to take place for the progenitors
with $M_{\rm MS} \gtaprx 20 M_{\odot}$ due to their large binding energy of 
the iron core, where
$M_{\rm MS}$ is the mass of the progenitor on the main-sequence.
For massive stars with $M_{\rm MS} \gtaprx 40M_{\odot}$, 
supernova explosion cannot take place and 
collapsed matter forms a black hole promptly (F99).
Several seconds after the core collapse 
a quasi-steady accretion disk 
is formed around the hole if the progenitor has sufficient
angular momentum (MacFadyen \& Woosley 1999), 
which is in a reasonable range of a progenitor model 
of Heger, Langer \& Woosley (2000).
Highly variable supercritical accretion with
0.01--0.1 $M_{\odot} \, \rm s^{-1}$ is 
maintained for approximately 10--20 s (MacFadyen \& Woosley 1999).
For such high accretion rates, however, 
the accretion disk has enough high temperature ($\gtaprx 10^{10}$ K)
for accreting gas to contain only protons and neutrons
(Popham, Fryer \& Woosley 1999). 
Therefore, the progenitors with $M_{\rm MS} \gtaprx 40M_{\odot}$ 
is irrelevant for p-process nucleosynthesis inside SSADs.

While for progenitors with $M_{\rm MS} = 20$--$40 M_{\odot}$ 
a neutron star is formed soon after the core collapse.
Supernova launches outward-going shock successfully
but some fraction of supernova ejecta falls back onto 
the nascent neutron star, 
which is transformed to a black hole via massive fallback (F99).
The accreting matter onto the hole forms a disk
with $10^{-4}$--$10^{-2} M_{\odot} \, \rm s^{-1}$ 
lasting for hundreds to thousands of seconds 
(MacFadyen et al.~2001).
After most of silicon layers of the progenitors has fallen back into the hole, 
the accretion rate is declined 
as $t^{-5/3}$ (Chevalier 1989; MacFadyen et al.~2001)
and compositions of the accreting matter are those of 
explosively burned oxygen/neon layers via supernova shock.
In a $20 M_{\odot}$ progenitor model (Hashimoto 1995), 
the position of $r_{\rm out}$ ($=3000 r_{\rm g}$) is far from the core and
located at $M_{r} \simeq 2.1 M_{\odot}$, or the oxygen/neon layer, 
where the peak temperature during the shock propagation 
is too low to proceed appreciable nuclear burning of s-process seeds
($ \ltaprx 2 \times 10^9$ K).
The material still leaves the s-process seeds even after the supernova explosion.
Appreciable amounts of the seeds can fall back 
onto the compact object through the accretion disk.
Accordingly, 
p-process nucleosynthesis operates inside a supercritical accretion disk
driven by supernova explosion of the
progenitors with $M_{\rm MS} = 20$--$40 M_{\odot}$.

\subsection{Initial Abundances of Fallback Matter}

Chemical composition of fallback matter depends on
masses of progenitors and fallback mechanisms.
Because p-process nucleosynthesis inside SSADs
needs s-process seeds, 
we set the compositions of the accreting gas 
at $r_{\rm out}$ to be
averaged compositions of oxygen/neon layers of the progenitors
of 10, 20, 30, and 40 $M_\odot$, whose models are refererred to 
as M10, M20, M30, and M40, respectively.
These abundances were calculated with a network
which contains 440 nuclei up to Bi. 
(Prantzos, Hashimoto \& Nomoto 1990a).
Although the progenitors with masses of $M_{\rm MS} < 20 M_{\rm \odot}$
are unlikely to be relevant for p-process inside a SSAD,
we also calculate abundances of p-nuclei for M10
because the abundance pattern is different from the others (Figure 1).
It should be emphasized that
these abundances were also adopted as the seed abundances 
in the p-process models of Rayet et al.~(1990, 1995).
Figure 1 shows the initial abundances of M10, M20, and M40, 
which are denoted by the dashed, solid, 
and thick-solid lines, respectively.
The initial abundances of M30 are in between those of M20 and M40.
The solar abundances are also depicted with the filled circles
(Anders \& Grevesse 1989).


Compositions of fallback matter 
are not those of the oxygen/neon layers of the progenitors but 
those of explosively burned layers because of supernova shock heating.
Abundances of p-nuclei produced inside SSADs 
depend on amounts of s-process seeds, as we can see later.
The explosively burned layers where the peak temperature of a shock wave 
is lower than 2$\times 10^9 \rm K$ still contain 
comparable s-process seeds to those before the explosion.
Accordingly, the initial compositions of the oxygen/neon layers of the progenitors 
are relevant for our p-process model inside SSADs
at least for fallback of the layers 
with relatively low peak temperature $\le 2\times 10^9 \rm K$.

The compositions of fallback material discussed above 
are possibly altered when large-scale matter mixing 
is significant in supernova explosion.
Such mixing is probably associated with fallback.
Hence, we shall discuss effects of matter mixing 
on p-process nucleosynthesis in \S 3.3.

\section{Nucleosynthesis inside SSADs}

\subsection{Abundance Distributions of P-Nuclei inside SSADs}

As fallback material accretes onto a central object, 
temperature and density of the gas increase (see equations (1) and (2)) 
and thus abundances of the accreting gas are changed 
through a sequence of nuclear reactions
from an initial composition described in \S 2.4. 
Using the nuclear reaction network
and the self-similar solution, 
we can follow the evolution of the chemical composition inside SSADs
during the infall toward the compact object
through post-processing calculations for M10, M20, M30, and M40.
The calculations are carried out from $r_{\rm out}$ to $r_{\rm in}$.
It should be noted that the nuclear energy generation is assumed 
to be much smaller than the viscous heating.

For disks with $\dot{m} \ltaprx 10^5$, 
the maximum temperature of disks is less than
$2 \times 10^9$ K which is the minimum temperature for the p-process
to proceed in the so called {\it p-process layers} (PPLs)
in Type II supernovae (Rayet et al.~1990, 1995).
Therefore, p-process nucleosynthesis does not operate
in the disks for such low accretion rates.
For higher accretion rates, the accreting gas
has higher temperature and the p-process can proceed significantly.
The abundance distributions of p-nuclei are similar 
for the disks with various accretion rates, 
while the production site of p-nuclei shifts 
toward the outer part of the disk as the accretion rates increase.
Thus, we show the results of $\dot{m} = 10^8$ as a representative case.

Figure 2 shows the abundance profiles, normalized with the solar abundances, 
 of representative p-nuclei for M20.
The abscissa is the radius of the fallback disk in units of 
$r_{\rm g}$.
The solid, thick-solid, dashed, thick-dashed, dotted, and thick-dotted 
lines indicate the abundances of
${}^{74}\rm Se$, ${}^{92}\rm Mo$, ${}^{138}\rm La$, 
${}^{152}\rm Gd$, ${}^{180}\rm Ta$, and ${}^{196}\rm Hg$, respectively.
At $r > 100~r_{\rm g}$, temperatures of the accreting gas
are too low to proceed appreciable synthesis of p-nuclei.
The abundances of ${}^{152}\rm Gd$ and ${}^{180}\rm Ta$ 
are the same as the initial values.
At $r \simeq 100~r_{\rm g}$ ($T \simeq 1.3\times 10^9$ K), 
${}^{180}\rm Ta$ increases via neutron capture of ${}^{179}\rm Ta$, 
contained in the fallback material initially.
As the gas moves to the inner part of the disk, 
the temperature increases and 
the heavy p-nuclei ($A \ge 150$), such as 
${}^{152}\rm Gd$, ${}^{180}\rm Ta$, and ${}^{196}\rm Hg$
are rapidly destroyed through $(\gamma, \rm n)$ reactions.
At $r \ltaprx 50~r_{\rm g}$ ($T \gtaprx 2.0\times 10^9$ K), however, 
these isotopes are drastically produced.
This is due to the photodisintegrations
of the neutron-rich seed elements
when the temperature of the accreting gas attains 
to $2-3\times 10^9$ K comparable to PPLs (Rayet et al.~1995).
Thus, the p-processes operate efficiently 
inside the accretion disk as in the PPLs.
The intermediate p-nuclei ($100 \le A < 150$) also enhance
via the p-process.
It should be noted that $t_{\rm dr} \simeq $ 0.1--1 $\rm s$ in this region 
is comparable to the time-scale of shock propagation 
through PPLs (Rayet et al.~1995).

For a region $r \ltaprx 30 r_{\rm g}$ ($T \gtaprx 2.8\times 10^9$ K), 
$(\gamma, \rm p)$ and $(\gamma, \alpha)$ reactions 
of the heavy and intermediate isotopes
dominate over $(\gamma, \rm n)$ reactions;
while for the light p-nuclei ($A < 100$), such as 
${}^{74}\rm Se$ and ${}^{92}\rm Mo$, 
$(\gamma, \rm n)$ reactions are still dominant processes.
The light p-nuclei are eventually depleted via 
$(\gamma, \rm p)$ and $(\gamma, \alpha)$ reactions
near $20~ r_{\rm g}$ ($T \sim 3.6\times 10^9$ K).
It is noted that the abundance profiles for all models are similar, 
while the more p-nuclei are produced the more massive are progenitors
because of the richness of s-process seeds (Figure 1).

\subsection{Ejected Masses of P-Nuclei from SSADs}

Many astrophysical objects involving an accretion disk are observed
to produce jets and/or winds (e.g., Livio 1999 and references therein).
Hence, as the accreting gas falls back onto the central object, 
some fraction of gas is possibly ejected from the disk
via jets and/or winds.
The processed material ranging from $r_{\rm in}$ to $r_{\rm ej}$ 
is supposed to be ejected.
Then we can estimate the averaged mass fraction
$\bar{X}_i$ of the {\itshape i}-th p-nucleus
ejected from the disks, 
\begin{equation}
 \bar{X}_i = \frac{2 \pi}{M(r_{\rm ej})}
\int_{r_{\rm in}}^{r_{\rm ej}} X_i(r) \, \Sigma(r) \, r \, dr, 
\end{equation}
where $\Sigma(r)$ is the column density of the disk.
Here $M(r_{\rm ej})$ is the mass of the disk from $r_{\rm in}$ to $r_{\rm ej}$
and the fraction of the ejected gas to the accreting gas 
is assumed to be constant in radius.
To compare the p-nuclei abundances ejected from the fallback disks
with the solar abundances, 
we calculate the overproduction factors (OPFs) $F_i$ of 35 p-nuclei
(Rayet et al.~1995):
\begin{equation}
 F_i = (\bar{X}_i/ X_{i, \, \odot})/F_0, 
\end{equation}
where $X_{i,\,\odot}$ is the mass fraction
in the solar system and 
$F_0 = \sum_{i = 1}^{35} (\bar{X}_i/X_{i,\,\odot})/35$.
In many accretion powered systems, 
matter ejection via jets and/or winds
could be originated from the inner region of the accretion disks.
Therefore, we set $r_{\rm ej}$ to be $100 r_{\rm g}$
(Junor et al.~1999).
The chemical composition of the ejected material
is assumed to be constant throughout a jet and/or a wind.
Abundances of p-nuclei increase due to
$\beta$-decays of their unstable parent nuclei after the freeze out.
However, abundances of most p-nuclei are not appreciably enhanced
after the $\beta$-decays (see \S 4.2).

Figures 3 and 4 show the OPFs of 35 p-nuclei
for M20 and M10, respectively.
It is emphasized that the abundance profiles
have similar features to those in Type II supernovae (Rayet et al.~1995).
The light p-nuclei, \nuc{Mo}{92, 94} and \nuc{Ru}{96,98}
are underproduced in our calculations.
This is due to the deficiency of their s-process seeds (Fuller \& Meyer 1995).
The intermediate mass p-nuclei, 
\nuc{In}{113}, \nuc{Sn}{115}, and \nuc{La}{138},
cannot be produced appreciably for M10 and M20.
It is noted that \nuc{In}{113} and \nuc{Sn}{115}
are largely enhanced through $\beta$-decays 
after ejection via jets and/or winds (see \S 4.2).
\nuc{Gd}{152} is underproduced in our calculations, 
while a large amount of \nuc{Gd}{152} is synthesized 
in the stellar s-process (Prantzos et al.~1990a).
\nuc{Er}{164} is also underproduced in our calculations.
The overproductions of \nuc{Se}{74} (M20) 
and \nuc{Os}{184} (M10) are attributed to
the ampleness of their s-process seeds.
The OPF of \nuc{Te}{120} is largely increased for M10 compared with that of M20.
The OPFs of Se, Kr, and Sr for M10 are smaller than those for M20
but those of p-nuclei with $A \ge 168$ are larger.
Such profiles of the OPFs are also responsible for 
initial abundance distributions of s-process seeds.
The OPFs for M30 and M40 are similar profiles to those for M20.
However, $F_0$ for M30 and M40 are larger than 
$F_0$ for M20 (Table 2).
This is because more massive progenitor has larger amounts
of s-process seeds (Figure 1).

Profiles of the OPFs ejected from SSADs 
depend on initial abundance distributions of s-process seeds.
The s-process seeds are mainly synthesized
during a helium core burning stage in massive stars. 
The s-process nucleosynthesis in massive stars
has been extensively investigated by many authors
(e.g., Prantzos et al.~1990a; K\"appeler et al.~1994; Rayet \& Hashimoto 2000); 
Uncertainties in some key nuclear reactions still preclude
precise estimation of s-process yields.

\subsection{Proton and Helium Contamination through Matter Mixing}

Several observations in SN 1987A support the occurrence of 
large-scale mixing in the ejecta. 
Newly synthesized \nuc{Ni}{56} near the collapsed core
is likely to be mixed up into the hydrogen envelope as indicated 
from observations in early phase;
X-ray and $\gamma$-ray light curves 
(e.g., Kumagai et al.~1989 and references therein) and
optical spectroscopy (e.g., Mitchell et al.~2001 and references therein).
Two-dimensional simulations of supernova explosions
(M\"uller, Fryxell \& Arnett 1991; Herant \& Benz 1992;
Nagataki, Shimizu \& Sato 1998)
indicate that the large-scale mixing is caused 
via Rayleigh-Taylor instability. 
A recent high-resolution simulation by Kifonidis et al.~(2000) 
shows that a large amount of $\rm {}^{4}He$ is mixed down 
near the iron core by the Rayleigh-Taylor instability.
Hydrogen is also shown to be mixed down to the core
(Hachisu et al.~1990).
Moreover, several analyses of observations in SN 1987A 
are also preferable to mixing of the outer hydrogen-rich envelope 
down to the inner metal-rich core;
the width of plateau-like peak of the bolometric light curves
(Shigeyama \& Nomoto 1990),
line profiles in late spectral evolution (Kozma \& Fransson 1998),
and the light curve for the first 4 months (Blinnikov et al.~2000).
Thus, appreciable amounts of proton and helium 
are likely to be mixed down to the inner core 
via the Rayleigh-Taylor instability in SN 1987A.
Certain fractions of proton and helium are accordingly mixed 
into the oxygen/neon layers.

Such mixing is probably generic in core-collapsed supernovae;
recent X-ray observations in Cas A (Douvion et al.~1999; Hughes et al.~2000),
emission line profiles of SN 1988A
(Spyromilio 1991) and SN 1993J (Spyromilio 1994), and
optical and infrared spectroscopies of SN 1995V (Fassia et al.~1998)
and SN 1998S (Fassia et al.~2001).

Moreover, recent investigation for anomalous composition 
in a companion of the black hole binary Nova Sco
suggests that 
the companion is polluted by material ejected in the supernova
accompanied with the formation of the hole
(Israelian et al.~1999).
The progenitor in this system is likely to 
experience substantial fallback ($\simeq$ a few $M_{\odot}$)
and matter mixing in supernova ejecta (Podsiadlowski et al.~2002).
Such fallback with mixing is also preferable to 
the large abundance of Zn observed in the very metal-poor stars
(Umeda \& Nomoto 2002).
Therefore, the fallback gas is expected to be contaminated 
with some fractions of proton and helium 
via the large-scale mixing.

To investigate effects of proton and helium contamination
of the fallback material on the p-process nucleosynthesis, 
we add an arbitrary fraction of proton or helium 
to the initial abundance of M20
and calculate the abundance distributions inside SSADs.
The initial proton abundance of the fallback matter highly depends on
large-scale mixing and fallback mechanism.
There exists no reliable estimation of the initial proton abundance.
Analyses of observations in SN 1987A indicate that the mass fraction of proton 
is 0.01 (Blinnikov et al.~2000) or up to 0.3 (Shigeyama \& Nomoto 1990).
Hence we examine the cases of the initial mass fraction of 
proton $X_{\rm p} = 0.01$, $0.1$, and $0.3$, 
which are referred as models M20P001, M20P01, and M20P03, respectively.
In these models, p-nuclei are produced via not only photodisintegration 
of s-process seeds but radiative proton capture of lighter nuclei.

Figure 5 shows the OPFs for M20P001 with $\dot{m} = 10^8$.
The most remarkable is an enhancement of \nuc{Mo}{92}.
Even if the fallback material contains a small fraction of proton, 
radiative proton captures as well as photodisintegrations 
are efficient.
The light p-nuclei, \nuc{Se}{74}, \nuc{Kr}{78}, and \nuc{Sr}{84}, are increased, 
while the light p-nuclei underproduced in M20,
such as \nuc{Mo}{94}, \nuc{Ru}{96}, and \nuc{Ru}{98}, are still deficient.
The intermediate mass p-nuclei, \nuc{Sn}{114}, \nuc{Sn}{115}, and
\nuc{La}{138} are underproduced as M20.
The OPFs of the heavy p-nuclei are decreased compared with M20
though abundances of these p-nuclei are not appreciably changed.
This is because the averaged OPF, $F_0$, is increased (equation (5)).
We also find that profiles of the OPFs weakly depend on 
$X_{\rm p}$ for $\dot{m} = 10^8$: 
the OPFs for M20P01 and M20P03 
are not significantly changed from those for M20P001.
The averaged OPFs, $F_0$, of M20P001, M20P01, and M20P03 
are larger than that of M20 and listed in Table 2.
For models with initial abundances of M10, M30, and M40 added by proton of 
$X_{\rm p} = 0.01$, $0.1$, and $0.3$, 
the OPFs are similar profiles to model with the initial abundance of 
20$M_{\odot}$ model added by the same fraction of proton
(i.e., same $X_{\rm p}$)
though the averaged OPFs, $F_0$, 
are significantly increased for models with massive progenitor (Table 2).

For $\dot{m} = 10^6$, 
the profiles of the OPFs change as protons are injected.
Appreciable amounts of the p-nuclei which are deficient in model M20
are produced in SSADs.
For $X_{\rm p} = 0.01$ (M20P001), 
\nuc{Mo}{92} and \nuc{La}{138} are overproduced.
The overproductions of \nuc{Mo}{92} and \nuc{La}{138} 
are more prominent in M20P01 (Figure 6) compared with M20P001.
When $X_{\rm p} = 0.3$ (model M20P03; Figure 7), 
\nuc{Ru}{96}, \nuc{Sn}{113}, and \nuc{Sn}{114} become abundant
because of very efficient proton captures.
However, the overall profile of the OPFs 
becomes worse compared with the solar one.
$F_0$ drastically increases compared with the cases 
of small proton inclusion;
$F_0$ are equal to 67.68, 41.84, and 2882, for M20P001, M20P01, and M20P03, 
respectively.
It should be noted that \nuc{Mo}{94}, \nuc{Ru}{98}, and \nuc{Sn}{115} 
are underproduced even if protons exist abundantly.

Concerning helium contamination of fallback material via large-scale mixing,
the OPFs are not appreciably changed from those of M20
even if a significant fraction of \nuc{He}{4}, up to 0.5, 
is added to the initial abundance of M20.
Hence, amounts of p-nuclei produced inside SSADs 
are independent of an amount of \nuc{He}{4} included in fallback material,
while $\alpha$ elements, such as \nuc{Ti}{44}, are enhanced.

\section{Discussion}

\subsection{Contribution of SSADs to Chemical Evolution of P-Nuclei}

We discuss contribution from SSADs to chemical evolution of p-nuclei
to estimate relative contribution from SSADs in respect to supernovae.
In this aim we average the overproduction factor $F_0$
of p-nuclei produced by supernovae or SSADs
using the initial mass function (IMF) $\xi(M)$. 
The IMF-weighted OPF is calculated as 
\begin{equation}
 \langle F_0 \rangle_{\rm IMF} = 
  \frac{ \int^{M_{\rm u}}_{M_{\rm l}} F_0(M) M_{\rm cont}(M)\xi(M) dM }
  { \int^{M_{\rm u}}_{M_{\rm l}} \xi(M) dM }
\end{equation}
where $M_{\rm l}$ and $M_{\rm u}$ 
are the lower and upper mass limits of progenitors and 
$M_{\rm cont}$ is the total mass (in units of $M_{\odot}$)
which contributes to the production of p-nuclei
via supernovae or SSADs.

We assume that the progenitors of $13 \le M_{\rm MS}/M_{\odot} \le 20$
can produce p-nuclei via supernova explosion (Rayet et al.~1995), 
while the progenitor with $20 \le M_{\rm MS}/M_{\odot} \le 40$ 
can yield p-nuclei through SSADs driven by massive fallback.
For supernovae and SSADs, 
$M_{\rm cont}$ is the mass of the PPLs, 
$M_{\rm PPL}$ (Rayet et al.~1995) and the ejected mass from a disk
via jets and/or winds, $M_{\rm ej}$, respectively.
Adopting $\xi(M) \propto M^{-2.3}$ proposed by Salpeter (1955), 
we can calculate $\langle F_0 \rangle_{\rm IMF}$ as 18.3 for supernovae.
Here we use the results of Rayet et al.~(1995) for
$M_{\rm PPL}$ and $F_{0}(M)$ (in their Tables 2 and 3).
A calculation is performed in a similar manner to Rayet et al.~(1995).

Next we estimate $\langle F_0 \rangle_{\rm IMF}$ for SSADs for the two cases:
no proton mixing and complete mixing.
In actual situations, 
uncertain fractions of fallback matter are affected by matter mixing.
Hence $\langle F_0 \rangle_{\rm IMF}$ for SSADs
is located between the two estimations.
Firstly we discuss the case in absence of matter mixing
and then those with complete mixing.

\subsubsection{The Case in The Absence of Matter Mixing}

To calculate $\langle F_0 \rangle_{\rm IMF}$ for SSADs, 
we need an estimation of the total fallback mass with 
s-process seeds, $M_{\rm seed}$. 
An amount of fallback material has been estimated by 
WW95 and F99 through
numerical simulations of core collapse and supernova explosion.
Amounts of fallback material $M_{\rm f}$ are presented in Table 3
for the fallback models of WW95 and F99.
To operate the p-process in SSADs, 
fallback material should contain the s-process seeds. 
Accordingly, to calculate $M_{\rm seed}$, 
besides the amount of fallback matter, 
both the abundance distribution of the progenitor 
and the peak temperature of the supernova shock are needed.
In Table 3, we also show the mass coordinate of the base of 
oxygen burning layers, $M_{\rm Ob}$ and
the mass coordinate where the peak temperature is
$2 \times 10^9$ K, $M_{\rm dest}$ 
for the progenitor and explosion models 
with masses of 15, 20, 25, and 40 $M_{\odot}$ (Hashimoto 1995).
All s-process seeds are assumed to be destroyed 
in the region of the mass coordinate $M_r \le M_{\rm dest}$ 
and thus disk p-process can operate only for
fallback material with $M_r > M_{\rm dest}$.
Hence we can estimate $M_{\rm seed}$, 
which are also presented in Table 3, for WW95 and F99.
For the fallback model of WW95, 
$M_{\rm seed}(30M_{\odot})$ is estimated as 2.41$M_{\odot}$
provided that $M_{\rm dest}$ for the 30$M_{\odot}$ progenitor
is equal to that of the 25$M_{\odot}$ progenitor.
For the fallback model of F99, 
we set simply 
$M_{\rm seed}(20M_{\odot})
=0.5[M_{\rm seed}(15M_{\odot})+ M_{\rm seed}(25M_{\odot})]$ 
and 
$M_{\rm seed}(30M_{\odot})
=\frac{1}{3}[2M_{\rm seed}(25M_{\odot})+ M_{\rm seed}(40M_{\odot})]$.

Moreover, some fractions of synthesized p-nuclei inside SSADs 
are ejected from the disks via jets and/or winds
as in many objects associated with an accretion disk (e.g., Livio 1999).
We assume that
the ratio $\epsilon$ of the mass ejection rate $\dot{M}_{\rm ej}$
to the mass accretion rate is constant in time,
or $\dot{M}_{\rm ej} = \epsilon \dot{M}$.
Hence the total ejected mass is also expressed as 
$M_{\rm ej} = \epsilon M_{\rm seed}$.
The ratio $\epsilon$ are 0.001--0.01 (Eggum, Coroniti \& Katz 1988)
and 0.03--0.1 (Kudoh, Matsumoto \& Shibata 1998)
estimated from numerical simulations
of jets originated from an accretion disk.
However, $\epsilon$ may be larger as suggested by observations
in X-ray binaries; SS433 (Kotani 1997) and GRS 1915+105 
(Fender \& Pooley 2000; Belloni et al.~2000).

Using the calculated values of $F_0$ for our SSAD models (Table 2), 
we evaluate $\langle F_0 \rangle_{\rm IMF}$ 
as 1.640 $\epsilon_{0.01}$ and 2.864 $\epsilon_{0.01}$
with the estimation of amounts of fallback matter of 
WW95 and F99, respectively, 
where $\epsilon_{0.01} = \epsilon/0.01$.
Thus $\langle F_0 \rangle_{\rm IMF}$ for SSADs 
are smaller than that for supernovae ($= 18.30$)
unless $\epsilon$ is as large as 0.1.
It should be noted that
$\langle F_0 \rangle_{\rm IMF}$ slightly decreases for a steep IMF slope
with $\xi(M) \propto M^{-2.7}$ (e.g., Kroupta 2002)
and also reduce by 30--50\% if $M_{\rm u} = 30M_{\odot}$.

\subsubsection{The Case with Complete Matter Mixing}

We proceed the case of fallback with complete proton mixing.
We assume matter contains proton of $X_{\rm p} = 0.01$
throughout fallback.
The proton mixing needs the hydrogen envelope of progenitors, 
which is decreased due to mass loss for massive progenitors
(e.g., Chiosi \& Maeder 1986).
The upper limit $M_{\rm u}$ possibly decreases for cases with proton mixing
and are taken to be $M_{\rm u} = 35 M_{\odot}$ (Fryer \& Kalogera 2001).
As in the absence of matter mixing, we calculate $\langle F_0 \rangle_{\rm IMF}$ 
as 5.027 $\epsilon_{0.01}$ and 9.583 $\epsilon_{0.01}$
for the fallback models 
of WW95 and F99, respectively.
Here we set 
$F_0(35M_{\odot})=0.5[F_0(30M_{\odot})+ F_0(40M_{\odot})]$.
The enhancement of $\langle F_0 \rangle_{\rm IMF}$ compared with
no proton mixing is responsible for the increase in
$F_0$ for the models with proton mixing (Table 2).
Hence, in the case with complete mixing, 
the SSADs-origin p-nuclei are likely to comparably contribute to the 
chemical evolution of p-nuclei as the supernovae-origin p-nuclei.

In particular, SSADs probably have significant contribution
to the \nuc{Mo}{92} evolution.
Overproduction of \nuc{Mo}{92} takes place
for any $\dot{m}$ (Figures 5-7).
Accordingly, the overproduction 
is realized inside SSADs throughout the duration in 
which a large fraction of 
fallback matter accretes onto a central compact object.
Even for the cases of small $X_{\rm p}$, 
\nuc{Mo}{92} is overproduced inside SSADs;
This is the case of $X_{\rm p} \gtaprx 0.005$.
Thus, contribution to the \nuc{Mo}{92} evolution
is considerable for a small amount of proton $\sim 0.01 M_{\odot}$ 
mixed into the fallback matter of 0.1 to several $M_{\odot}$.

On the other hand, \nuc{Ru}{96} and \nuc{La}{138}
are overproduced only for the case of $\dot{m} = 10^6$.
The accretion rate of fallback matter is declined to
relatively small accretion rates $\dot{m} \sim 10^6$
at several months after the explosion (Mineshige et al.~1997)
and large amounts of fallback matter have already accreted onto a central 
remnant. Therefore, only a small amount of fallback matter 
are processed inside SSADs, which are enable to 
overproduce \nuc{Ru}{96} and \nuc{La}{138}.

\subsection{Abundance Change of P-Nuclei through Jets and Winds}

In SSADs, p-nuclei are mainly produced in the region 20--70$~r_{\rm g}$
for $\dot{m} = 10^8$ (Figure 2) where an accreting gas is mainly ejected via winds.
Parent nuclei which decay into p-nuclei through $\beta$-decays 
are also expelled via winds and thus enhance amounts of p-nuclei.
Taking into account the $\beta$-decays after freeze out,  
we estimate abundances of p-nuclei ejected from a SSAD.
We find that p-nuclei other than 
\nuc{Mo}{92}, \nuc{Mo}{94}, \nuc{In}{113}, and \nuc{Sn}{115} 
are not appreciably increased via $\beta$-decays of their parent nuclei;
Their ratio after to before $\beta$-decays
are 1.100, 1.215, 13.77, and 18.33, respectively, for M20.
Here all the parent nuclei has been assumed to decay to their daughter p-nuclei.
When proton is abundantly mixed into fallback material, 
the abundances of \nuc{Mo}{92} and \nuc{Mo}{94} are not 
significantly changed 
but \nuc{In}{113} and \nuc{Sn}{115} are largely enhanced.
For M20P03 ($X_{\rm p} = 0.3$), 
the ratio of the abundances of \nuc{In}{113}, and \nuc{Sn}{115}
after to before $\beta$-decays become 9.797 and 42.41, respectively.
However, the averaged OPFs, $F_0$, are not appreciably changed 
because of small OPFs of \nuc{In}{113} and \nuc{Sn}{115} up to $\sim$ 0.07
before $\beta$-decays (Figures 3--7).
Thus, $\beta$-decays after freeze out has only minor effect on estimation
of $\langle F_0 \rangle_{\rm IMF}$ discussed in \S 4.1.

\subsection{Neutron Star Kick }

If a compact object breaks out a supernova remnant, 
matter fallback onto the object ceases.
Many observations reveal that pulsars have larger space velocities
(200--1000 $\rm km\,s^{-1}$) than those of their progenitor
(e.g., Lai et al.~2001 and references therein).
Such high velocities are invoked to kicks of nascent neutron stars 
via asymmetric supernova explosions.
The supercritical accretion ($\dot{m} > 10^5$)
investigated in the present paper is realized until about 4 years 
after an explosion (Mineshige et al.~1997).
The expansion velocity of the supernova remnant is much larger 
than the typical kick velocity of pulsars.
The material therefore continues to fall back 
during the supercritical accretion of our interest.
However, the later expansion of the remnant is decelerated 
by ambient material from progenitor winds.
The pulsar eventually breaks through the supernova remnant 
and matter fallback onto it ceases.

\section{Concluding Remarks}

We have investigated the p-process nucleosynthesis 
inside SSADs
around a compact object of $1.4 M_{\odot}$. 
Supernova explosion of 20--40$M_{\odot}$ progenitors
is probably responsible for the p-process nucleosynthesis
inside SSADs.
The chemical compositions of the accreting gas far from 
the central object have been assumed to be 
those of the oxygen/neon layers of the progenitor in
the model of 10, 20, 30, and 40$M_\odot$ stars.
It is found that when $\dot{m} > 10^5$
the temperature of the accreting gas attains to $2-3 \times 10^9$ K
and appreciable nuclear reactions take place to produce the p-nuclei.
We have also estimated the OPFs of the p-nuclei
in the ejected material from the accretion disks via jets and/or winds 
to compare the abundances of the ejected matter with the solar abundances.
The resultant p-nuclei profiles have similar features to
those for Type II supernovae (Rayet at al. 1995).
The light and intermediate mass p-nuclei such as
\nuc{Mo}{92}, \nuc{Mo}{94}, \nuc{Ru}{96}, \nuc{Ru}{98}, 
\nuc{Sn}{114}, and \nuc{La}{138}
are underproduced.
The amounts of p-nuclei and profiles of the OPFs ejected from SSADs 
depend on initial abundance distributions of s-process seeds.

Moreover, for investigating the effects of proton and helium contamination
via large-scale mixing in a supernova explosion
of fallback material on the p-process nucleosynthesis, 
we have added some fractions of proton or helium 
to the initial composition for the cases of 
10, 20, 30, and 40$M_{\odot}$ progenitors
and calculated the OPFs of the p-nuclei in the ejected material.
It has been found that significant fractions of \nuc{He}{4} inclusion 
cause small changes in the OPFs.
Even for a small amount of proton included in fallback material, 
p-nuclei are synthesized through not only photodisintegrations
but radiative proton captures.
The most prominent is an enhancement of \nuc{Mo}{92}.
\nuc{La}{138} is also overproduced for $\dot{m} = 10^6$.
If fallback matter includes abundant proton, 
appreciable amounts of the p-nuclei, such as 
\nuc{Mo}{92}, \nuc{Ru}{96}, \nuc{Sn}{113}, and \nuc{Sn}{114}, 
which are deficient in Type II supernovae, can be produced for $\dot{m} = 10^6$, 
though the overall profiles of the OPFs becomes worse
compared with the solar ones.

We have discussed the contribution from the p-process nucleosynthesis 
in SSADs to galactic chemical evolution of p-nuclei.
We conclude that p-process in SSADs comparably contributes
to chemical evolution of p-nuclei as in supernovae
if several percents of fallback matter are ejected from SSADs.
In particular the p-process in SSADs may be important
for the chemical evolution of \nuc{Mo}{92} 
and possibly compensates the underproduction in Type II supernovae.
Contribution from SSADs to the chemical evolution of
\nuc{Ru}{96} and \nuc{La}{138} is unlikely to be significant.

Numerical simulations of supernova-driven jets present 
the peculiar nucleosynthesis of the alpha-rich freeze out 
(Nagataki et al.~1997): The p-process will be affected by jets as SN1987A.
Some gamma-ray bursts are likely to be caused by
aspherical jet-induced supernova explosions
(MacFadyen \& Woosley 1999; MacFadyen et al.~2001).
Numerical calculations indicate that in such explosions
a SSAD is formed around a new-born compact object 
and a substantial amount of fallback ($\simeq 0.2 M_{\odot}$) 
followed by mixing takes place
(Fryer \& Heger 2000; H\"oflich, Khokhlov \& Wang 2001).
Hence some gamma-ray bursts possibly accompany 
with the synthesis of p-nuclei;
Our SSAD p-process model would be responsible for the synthesis of 
p-nuclei in these bursts.

\acknowledgments

We are grateful to Dr.~Takashi Yoshida for his helpful processing 
nuclear data and useful discussion. 
We also thank the anonymous referee for useful comments.




\begin{deluxetable} {lclclc}
\tablewidth{23pc} 
\tablecaption{Elements included in the nuclear reaction network.}
\tablehead{
\colhead{elements} & \colhead{A} & \colhead{elements} &
\colhead{elements} & \colhead{A} }

\startdata 
 H &   1--  3 & Cu &  56-- 71 & La & 121--139 \nl 
He &   3--  6 & Zn &  57-- 74 & Ce & 126--142 \nl 
Li &   6--  8 & Ga &  60-- 77 & Pr & 127--143 \nl 
Be &   7-- 10 & Ge &  66-- 76 & Nd & 132--150 \nl 
 B &   8-- 12 & As &  70-- 77 & Pm & 133--151 \nl 
 C &  11-- 14 & Se &  71-- 82 & Sm & 136--154 \nl 
 N &  12-- 15 & Br &  74-- 83 & Eu & 137--155 \nl 
 O &  14-- 20 & Kr &  75-- 86 & Gd & 140--160 \nl 
 F &  17-- 22 & Rb &  78-- 87 & Tb & 143--161 \nl 
Ne &  17-- 24 & Sr &  80-- 88 & Dy & 146--164 \nl 
Na &  20-- 27 &  Y &  82-- 92 & Ho & 149--165 \nl 
Mg &  20-- 29 & Zr &  84-- 96 & Er & 151--170 \nl 
Al &  22-- 31 & Nb &  86-- 97 & Tm & 152--171 \nl 
Si &  24-- 34 & Mo &  88--100 & Yb & 154--176 \nl 
 P &  27-- 38 & Tc &  90--101 & Lu & 156--177 \nl 
 S &  28-- 42 & Ru &  92--104 & Hf & 158--180 \nl 
Cl &  31-- 45 & Rh &  94--105 & Ta & 164--181 \nl 
Ar &  32-- 48 & Pd &  96--110 &  W & 166--186 \nl 
 K &  35-- 49 & Ag &  98--111 & Re & 168--187 \nl 
Ca &  36-- 50 & Cd & 100--116 & Os & 170--192 \nl 
Sc &  39-- 51 & In & 103--117 & Ir & 172--193 \nl 
Ti &  40-- 53 & Sn & 105--124 & Pt & 174--198 \nl 
 V &  43-- 55 & Sb & 106--125 & Au & 178--200 \nl 
Cr &  44-- 58 & Te & 108--130 & Hg & 180--204 \nl 
Mn &  46-- 64 &  I & 114--131 & Tl & 186--205 \nl 
Fe &  47-- 65 & Xe & 116--136 & Pb & 190--208 \nl 
Co &  50-- 66 & Cs & 117--137 & Bi & 192--209 \nl 
Ni &  51-- 68 & Ba & 120--138 & \nodata & \nodata
\enddata 
\end{deluxetable}

\begin{deluxetable}{ccccc}
\tablewidth{22pc}
\tablecaption{Overproduction factor $F_0$ ejected from SSADs
 for the progenitors of 10, 20, 30, and 40 $M_{\odot}$ 
 and for various values of proton contamination $X_{\rm p}$.}
\tablehead{
\colhead{$M_{\rm MS}$} & \multicolumn{4}{c}{$F_0$\tablenotemark{a}} \\ \cline{2-5}
 ($M_{\odot}$) & $X_{\rm p} = 0$ & $X_{\rm p} = 0.01$ & 
   $X_{\rm p} = 0.1$ & $X_{\rm p} = 0.3$
}
\startdata 
 10 &  18.15  &  26.73  &  21.40  &  18.03 \nl
 20 &  49.64  &  161.1  &  88.46  &  89.03 \nl
 30 &  102.0  &  461.8  &  259.1  &  235.9 \nl
 40 &  138.3  &  698.5  &  422.3  &  332.2
\tablenotetext{a}{All $F_0$ are calculated for the case of $\dot{m} = 10^8$.}
\enddata
\end{deluxetable}

\begin{deluxetable}{ccccccccc}
\tablewidth{31pc}
\tablecaption{Various masses of progenitors and fallback matter}
\tablehead{
  & & &
  \multicolumn{3}{c}{WW95 model} &
  \multicolumn{3}{c}{F99 model} \\
  \colhead{$M_{\rm MS}$\tablenotemark{a}} &
  \colhead{$M_{\rm Ob}$\tablenotemark{b}} & 
  \colhead{$M_{\rm dest}$\tablenotemark{c}} &
  \colhead{$M_{\rm f}$\tablenotemark{d}} & 
  \colhead{$M_{\rm rem}$\tablenotemark{e}} &
  \colhead{$M_{\rm seed}$\tablenotemark{f}}  & 
  \colhead{$M_{\rm f}$\tablenotemark{d}} &
  \colhead{$M_{\rm rem}$\tablenotemark{e}} &
  \colhead{$M_{\rm seed}$\tablenotemark{f}}
}
\startdata 
 15  &  1.37   &  1.55   &  0.11  &  1.43  &   0     &  0.2    &  1.52  &   0    \nl 
 20  &  1.67   &  1.8    &  0.32  &  2.06  &  0.26   & \nodata & \nodata & \nodata \nl 
 25  &  1.48   &  1.8    &  0.29  &  2.07  &  0.27   &  3.8    &  5.58 &  2.0   \nl 
 30  & \nodata & \nodata &  2.41  &  4.24  & \nodata & \nodata & \nodata & \nodata \nl 
 40  &  2.36   &  3.5    &  8.36  &  10.34 &  4.84   &  10.3   &  11.01 &  6.8 \nl 
\enddata
\tablecomments{All masses are in units of $M_{\odot}$.}

\tablenotetext{a}{Progenitor mass on the main-sequence.}
\tablenotetext{b}{Location of the base of oxygen burning shell from
 Hashimoto (1995).}
\tablenotetext{c}{Location where the peak temperature of supernova
 shock is equal to 2 $\times 10^9$ K from Hashimoto (1995).}
\tablenotetext{d}{Fallback mass.}
\tablenotetext{e}{Remnant mass after fallback.}
\tablenotetext{f}{Estimated fallback mass with s-process seeds for p-nuclei.}
\end{deluxetable}

\newpage


 \begin{figure}
 \plotone{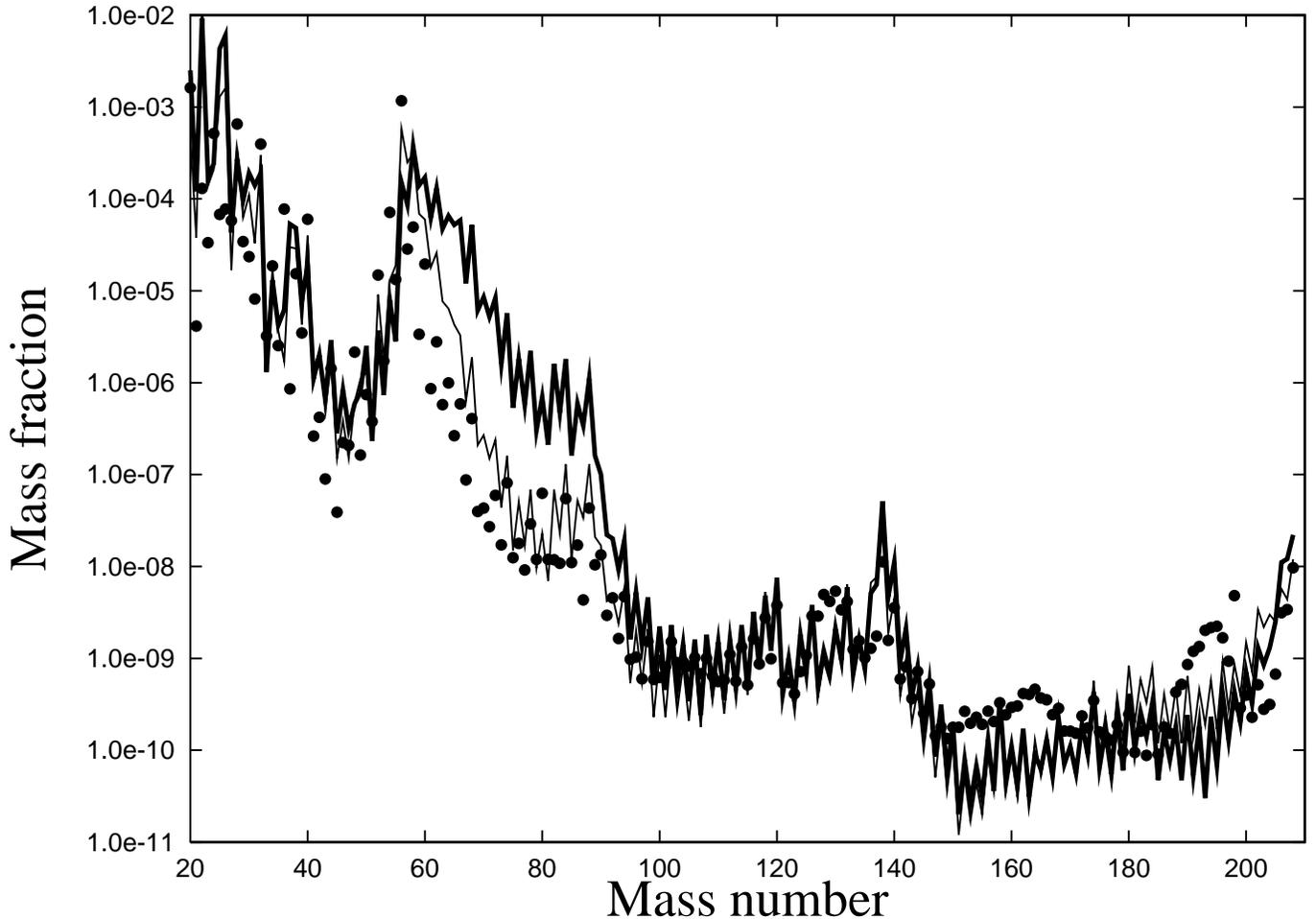}
 \caption{ The initial abundances of models M10 (dashed line) 
  M20 (solid line), and M40 (thick-solid line).
  The filled circles are the solar abundance (Anders \& Grevesse 1989).
  }
 \end{figure}

 \begin{figure}
 \plotone{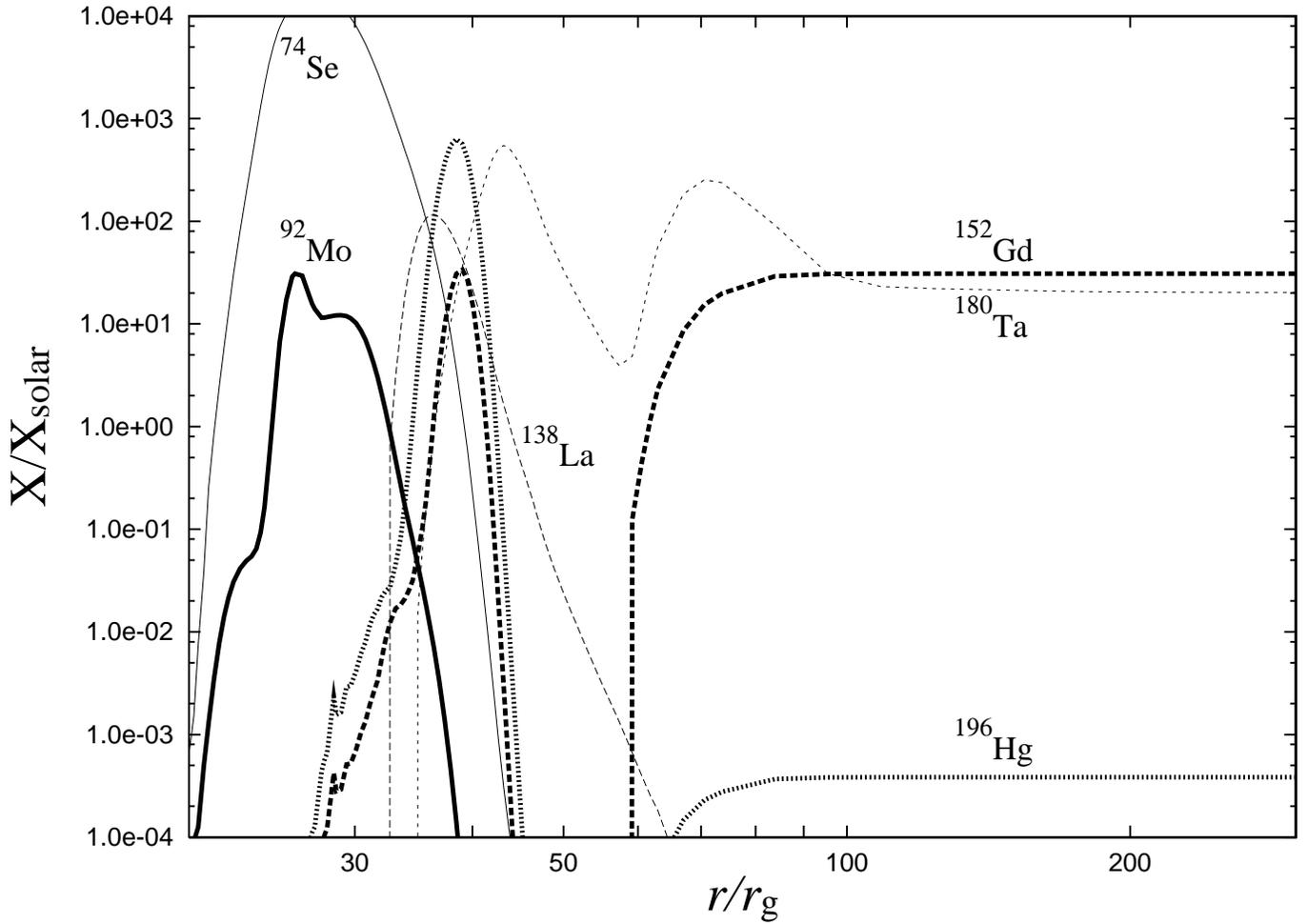}
 \caption{ The abundance profiles of representative p-nuclei 
  inside a SSAD for M20. 
  The abscissa is the radius of the fallback disk in units of
  the Schwarzschild radius.
  The solid, thick-solid, dashed, thick-dashed, dotted, and thick-dotted 
  lines represent 
  ${}^{74}\rm Se$, ${}^{92}\rm Mo$, ${}^{138}\rm La$, 
  ${}^{152}\rm Gd$, ${}^{180}\rm Ta$, and ${}^{196}\rm Hg$, respectively.
  }
 \end{figure}

 \begin{figure}
 \plotone{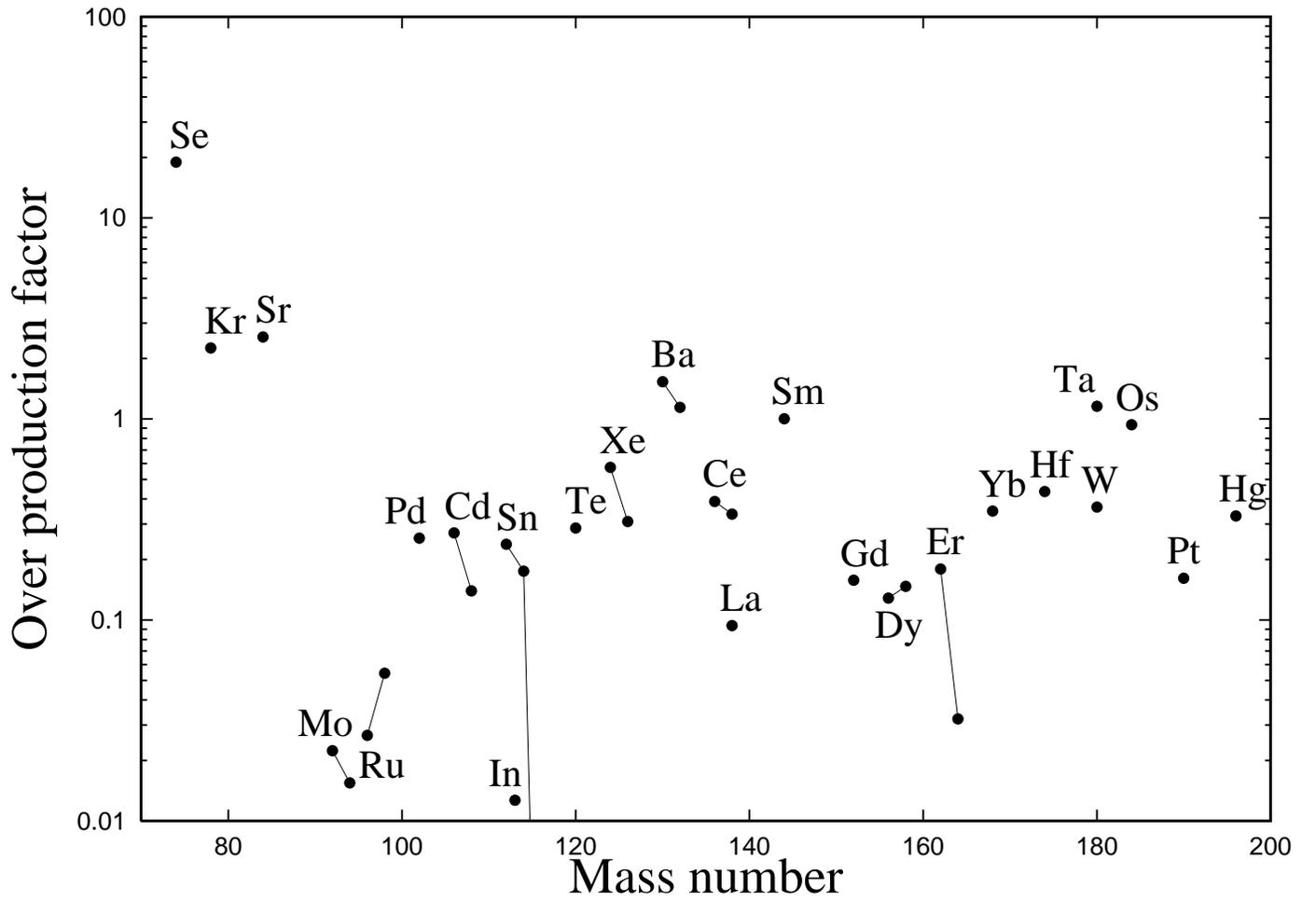}
 \caption{ OPFs of 35 p-nuclei for M20 with $\dot{m} = 10^8$.
  $F_0$ is equal to 49.64.}
 \end{figure}

 \begin{figure}
 \plotone{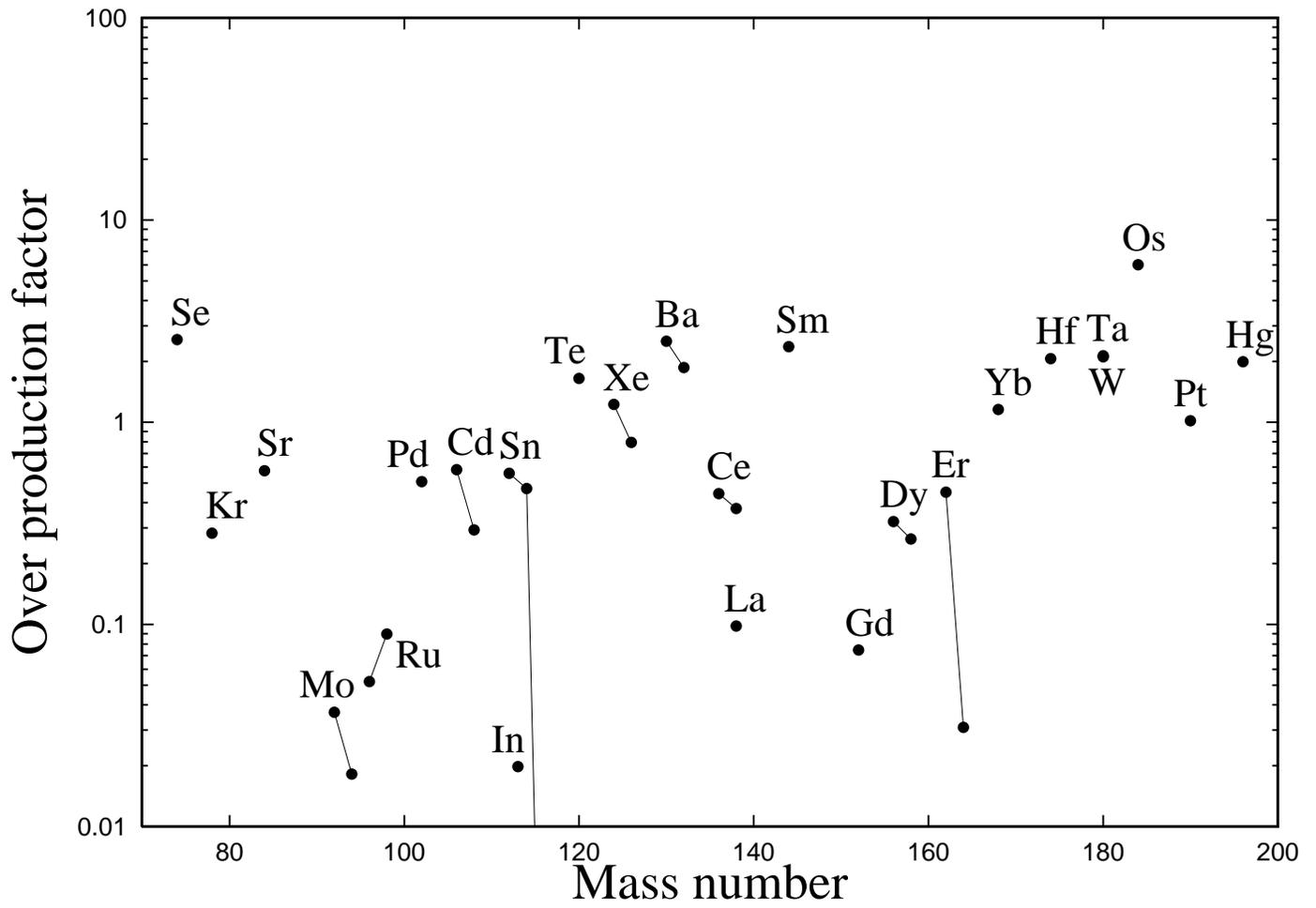}
 \caption{ Same as figure 3, but for M10. 
  $F_0$ is equal to 18.51.}
 \end{figure}

 \begin{figure}
 \plotone{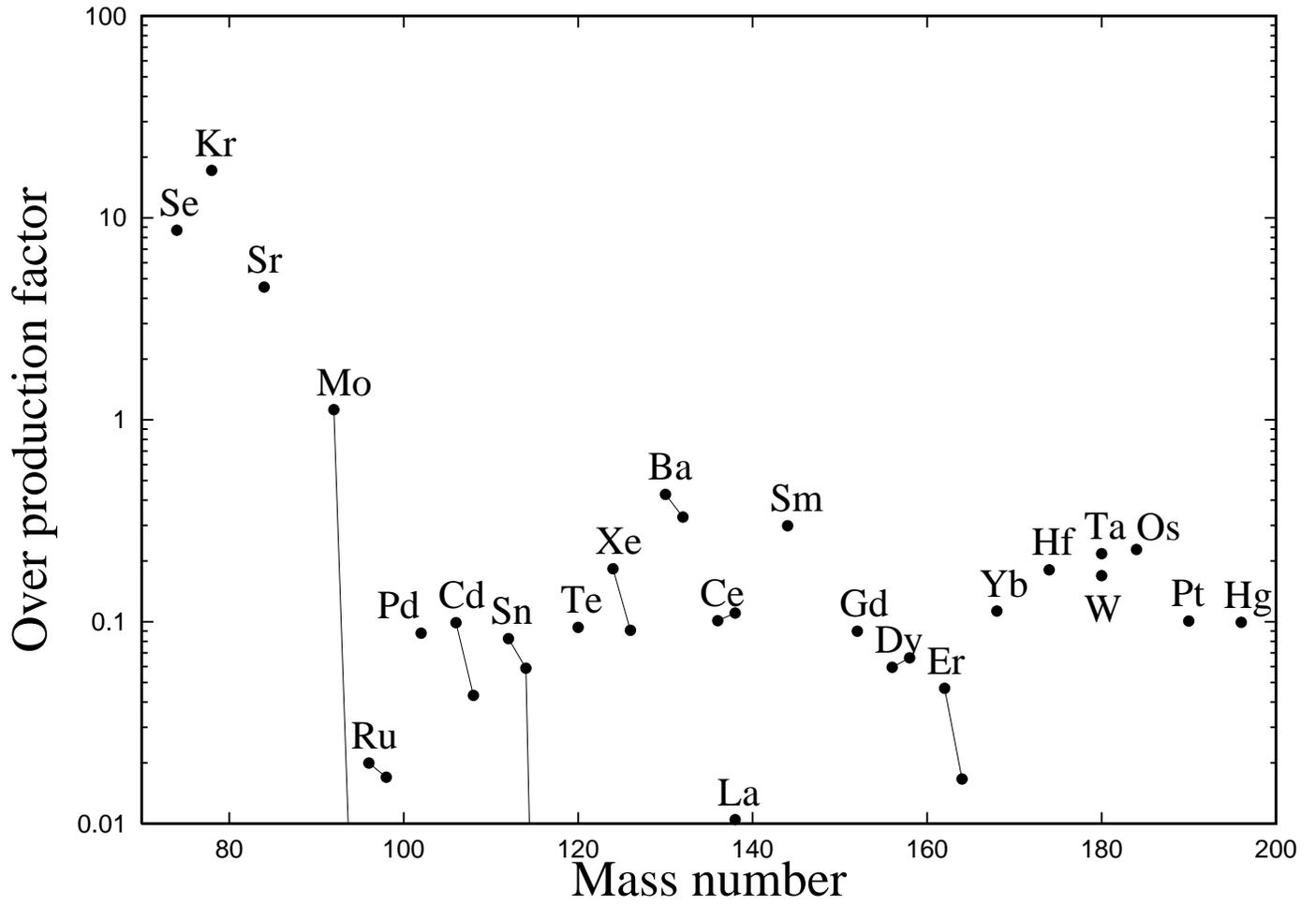}
 \caption{ Same as figure 3, but for M20P001. 
 $F_0$ is equal to 157.7.}
 \end{figure}

 \begin{figure}
 \plotone{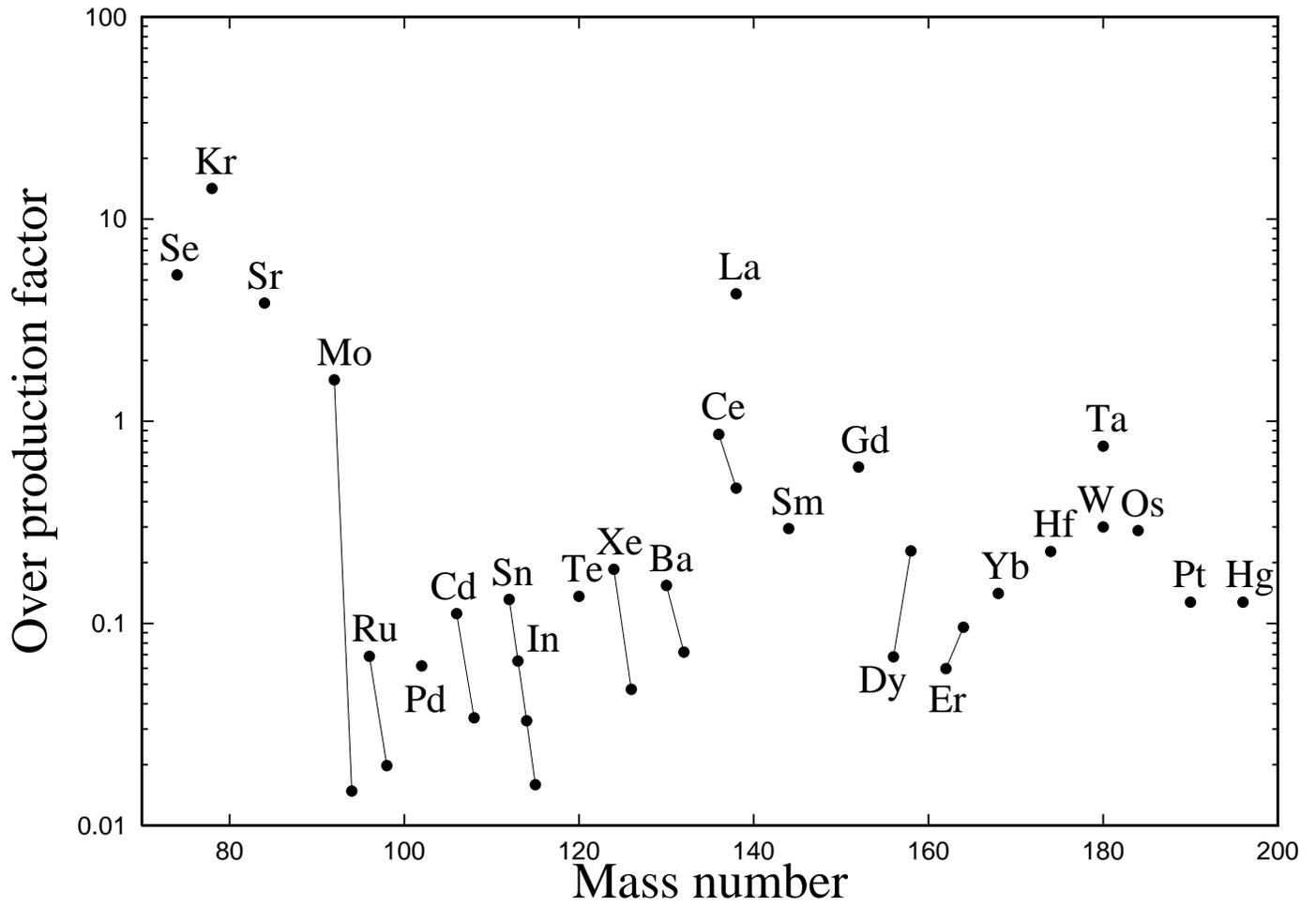}
 \caption{ OPFs of 35 p-nuclei for M20P01 with $\dot{m} = 10^6$.
 $F_0$ is equal to 41.84}
 \end{figure}

 \begin{figure}
 \plotone{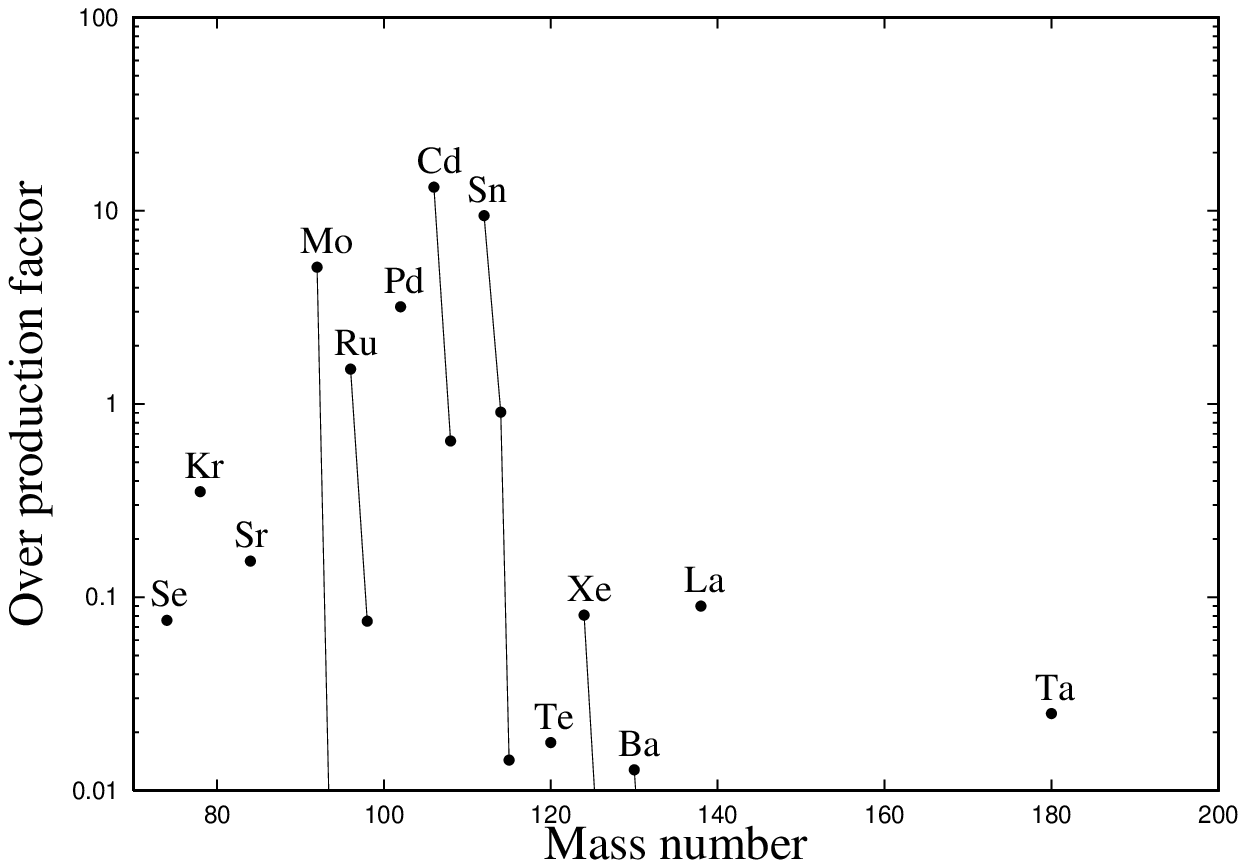}
 \caption{ Same as figure 6, but for model M20P03. 
 $F_0$ is equal to 2882}

 \end{figure}


\begin{thebibliography}{}

\bibitem{1}
Anders, E., \& Grevesse, N. 1989, \gca, 53, 197

 \bibitem{5}
Arnould, M. 1976, \aap, 46, 117

\bibitem{6} 
Arnould, M., Rayet, M., Hashimoto, M. 1998, 
in Tours Symposium on Nuclear Physics III,
ed. M. Arnould et al. (New York: AIP), 626

 \bibitem{7}
Audi, G., \& Wapstra, A. H. 1995, \nphysa, 595, 409

 \bibitem{a12}
Belloni, T., Migliari, S., \& Fender, R.\ P.\ 2000, \aap, 358, L29 

 \bibitem{9}
Blinnikov, S., Lundqvist, P., Bartunov, O., Nomoto, K., \& Iwamoto, K. 2000, \apj, 532, 1132

 \bibitem{10}
Chevalier, R.\ A.\ 1989, \apj, 346, 847 

 \bibitem{a5}
Chiosi, C., Maeder, A. 1986, \araa, 24, 329.

 \bibitem{11}
Colgate, S.\ A.\ 1971, \apj, 163, 221 

 \bibitem{12}
Costa, V., Rayet, M., Zappal\`a, R.~A., \& Arnould, M. 2000, \aap, 358, 67

 \bibitem{13}
Douvion, T., Lagage, P.~O., \& Cesarsky, C.~J. 1999, \aap, 352, L111

 \bibitem{a8}
Eggum, G.\ E., Coroniti, F.\ V., \& Katz, J.\ I.\ 1988, \apj, 330, 142 

 \bibitem{15}
Fassia, A., Meikle, W.~P.~S., Geballe, T.~R., Walton, N.~A., Pollacco, D.~L., Rutten, R.~G.~M., \& Tinney, C. 1998, \mnras, 299, 150

 \bibitem{17}
Fassia, A., et al. 2001, \mnras, 325, 907

 \bibitem{a9}
Fender, R.\ P., \& Pooley, G.\ G.\ 2000, \mnras, 318, L1 

 \bibitem{a2}
Fryer, C. L. 1999, \apj, 522, 413

 \bibitem{a13}
Fryer, C. L., Colgate, S. A., \& Pinto, P. A. 1999, \apj, 511, 885

 \bibitem{18}
Fryer, C. L., \& Heger, A. 2000, \apj, 541, 1033

 \bibitem{a6}
Fryer, C. L., Kalogera, V. 2001, \apj, 554, 548

 \bibitem{19}
Fujimoto, S., Arai, K., Matsuba, R., Hashimoto, M., Koike, O., \& Mineshige, S. 2001, PASJ, 53, 509

 \bibitem{20} 
Fuller, G.~M., Fowler, W.~A., \& Newman, M.~J. 1980, \apjs, 42, 447

 \bibitem{21} 
 -----------. 1982a, \apjs, 48, 279

 \bibitem{22} 
 -----------. 1982b, \apj, 252, 715

 \bibitem{23}
Fuller, G.~M., \& Meyer, B.~S. 1995, \apj, 453, 792

 \bibitem{24}
Goriely, S., Arnould, M., Borzov, I., \& Rayet, M. 2001, \aap, 375, 35

 \bibitem{25}
Goriely, S., Jose, J., Hernanz, M., Rayet, M., \& Arnould, M. 2002, \aap, 383, L27

 \bibitem{26}
Hachisu, I., Matsuda, T., Nomoto, K., \& Shigeyama, T.\ 1990, \apjl, 358, L57

 \bibitem{27}
Hashimoto, M.\ 
 1995, Prog.\ Theor.\ Phys., 94, 663

 \bibitem{a16}
Hashimoto, M., Arai, K. 1985, Phys.\ Rep.\ Kumamoto Univ., 7, 1

 \bibitem{a3}
Heger, A., Langer, N., \& Woosley, S. E. 2000, \apj, 528, 368

 \bibitem{28}
Herant, M., \& Benz, W. 1992, \apj, 387, 294

 \bibitem{29}
Hoffman, R.~D., Woosley, S.~E., Fuller, G.~M., \& Meyer, B.~S. 1996, \apj, 460, 478

 \bibitem{30}
Horiguchi, T., Tachibana, T., \& Katakura, J. 1996, 
Chart of The Nuclides, Nuclear Data Center, 
Japan Atomic Energy Research Institute, Ibaraki

 \bibitem{31}
Howard, W. M., Meyer, B. S., Woosley, S. E. 1991, \apjl, 373, L5

 \bibitem{a7}
Hawley, J.\ F.\ 2000, \apj, 528, 462

 \bibitem{32}
H\"oflich, P., Khokhlov, A., \& Wang, L. 2001, 
in 20th Texas Symposium on Relativistic Astrophysics,
ed. J. C. Wheeler \& H. Martel (New York:AIP), 459

 \bibitem{34}
Hughes, J. P., Rakowski, C. E., Burrows, D. N., \& Slane, P. O. 2000, \apjl, 528, L109

 \bibitem{35}
Israerian, G., Rebolo, R., Basrl, G., Casares, J., \& Martin, E.\ L.\
1999, \nat, 401, 142

 \bibitem{36}
Jaeger, M., Kunz, R., Mayer, A., Hammer, J.W., Staudt, G., Kratz, K.-L., \& Pfeiffer, B. 2001, \prl, 87, 202501

 \bibitem{37}
Junor, W., Biretta, J.\ A., \& Livio, M.\ 1999, \nat, 401, 891

 \bibitem{38}
K\"appeler, F., et al. 1994, \apj, 437, 396

 \bibitem{39}
Kifonidis, K., Plewa, T., Janka, H.-Th., \& M\"{u}ller, E.\ 
 2000, \apjl, 531, L123

 \bibitem{40}
Koike, O., Hashimoto, M., Arai, K., \& Wanajo, S.\ 1999, \aap, 342, 464

 \bibitem{a10}
Kotani, T.\ 1997, PhD thesis, The University Tokyo

 \bibitem{a11}
Kudoh, T., Matsumoto, R., \& Shibata, K.\ 1998, \apj, 508, 186

 \bibitem{41}
Kozma, C., \& Fransson, C. 1998, \apj, 497, 431

 \bibitem{a15}
Kroupta, P. 2002, Science, 295, 5552

 \bibitem{42}
Kumagai, S., Shigeyama, T., Nomoto, K., Itoh, M., Nishimura, J., \& Tsuruta, S. 1989, \apj, 345, 412

 \bibitem{43}
Lai, D., Chernoff, D. F., \& Cordes, J. M. 2001, \apj, 549, 1111

 \bibitem{44}
Livio, M. 1999, \physrep, 311, 225

 \bibitem{45}
MacFadyen, A. I., \& Woosley, S. E. 1999, \apj, 524, 262

 \bibitem{46}
MacFadyen, A. I., Woosley, S. E., \& Heger, A. 2001, \apj, 550, 410

 \bibitem{47}
Mineshige, S., Nomura, H., Hirose, M., Nomoto, K., \& Suzuki, T. 1997, \apj, 489, 227

 \bibitem{48}
Mitchell, R.~C., Baron, E., Branch, D., Lundqvist, P., Blinnikov, S., Hauschildt, P.~H., \& Pun, C.~S.~J. 2001, \apj, 556, 979

 \bibitem{49}
M\"uller, E., Fryxell, B., \& Arnett, D. 1991, \aap, 251, 505

 \bibitem{50}
Nagataki, S., Shimizu, T.~M., \& Sato, K. 1998, \apj, 495, 413

 \bibitem{a14}
Nagataki, S., Hashimoto, M., Sato, K., Yamada, S. 1997, \apj, 486, 1026

 \bibitem{51}
Podsiadlowski, P., Nomoto, K., Maeda, K., Nakamura, T., Mazzali, P., \& Schmidt, B. 2002, \apj, 567, 491

 \bibitem{a4}
Popham, R., Woosley, S. E., \& Fryer, C. 1999, \apj, 518, 356


 \bibitem{53}
Prantzos, N., Hashimoto, M., \& Nomoto, K. 1990a, \aap, 234, 211

 \bibitem{54}
Prantzos, N., Hashimoto, M., Rayet, M., \& Arnould, M. 1990b, \aap, 238, 455

 \bibitem{55} 
Rauscher, T., \& Thielemann, F.-K. 2000, At. Data Nucl. Data Tables, 75, 1

 \bibitem{56} 
 -----------. 2001, At. Data Nucl. Data Tables, 79, 47

 \bibitem{57}
Rayet, M., Arnould, M., Hashimoto, M., Prantzos, N., \& Nomoto, K. 1995, \aap, 298, 517

 \bibitem{58}
Rayet, M., \& Hashimoto, M. 2000, \aap, 354, 740

 \bibitem{59}
Rayet, M., Prantzos, N., \& Arnould, M. 1990, \aap, 227, 271

 \bibitem{a1}
Salpeter, E.E. 1955, \apj, 121, 161

 \bibitem{60}
Schatz, H., et al. 1998, \physrep, 294, 167

 \bibitem{61}
Shigeyama, T., \& Nomoto, K.\ 1990, \apj, 360, 242

 \bibitem{62} 
Spyromilio, J. 1991, \mnras, 253, 25

 \bibitem{63} 
 -----------. 1994, \mnras, 266, 61


 \bibitem{65}
Timmes, F.~X. 1999, \apjs, 124, 241

 \bibitem{67}
Umeda, H., \& Nomoto, K. 2002, \apj, 565, 385

 \bibitem{69}
Woosley, S. E., \& Howard, W.~M. 1978, \apjs, 36, 285

 \bibitem{a18}
Woosley, S. E., Hartmann, D. H., Hoffman, R. D.,\& Haxton, W. C. 1990, \apj, 356, 272

 \bibitem{70}
Woosley, S.\ E., \& Weaver, T.\ A.\ 1995, \apjs, 101, 181 (WW95)


\end{thebibliography}
\end{document}